\documentclass[reprint,aps,pra,superscriptaddress,showpacs,amsmath,amssymb,floatfix]{revtex4-1}

\usepackage{txfonts}

\allowdisplaybreaks[4]

\usepackage{graphicx}
\usepackage{txfonts}
\usepackage{subfigure}
\usepackage{braket}
\usepackage{bm}
\usepackage{multirow}

\usepackage{color}

\graphicspath{{./Figures/}}

\def\approxprop{%
  \def\p{%
    \setbox0=\vbox{\hbox{$\propto$}}%
    \ht0=0.6ex \box0 }%
  \def\s{%
    \vbox{\hbox{$\sim$}}%
  }%
  \mathrel{\raisebox{0.7ex}{%
      \mbox{$\underset{\s}{\p}$}%
    }}%
}

\newcommand{\UTokyo}{
Department of Applied Physics, School of Engineering, The University of Tokyo, \\
7-3-1 Hongo, Bunkyo-ku, Tokyo 113-8656, Japan}
\newcommand{\RQC}{
Optical Quantum Computing Research Team, RIKEN Center for Quantum Computing, \\
 2-1, Hirosawa, Wako, 351-0198, Tokyo, Japan}
\newcommand{\Xanadu}{
Xanadu, ON, M5G 2C8, Canada}

\begin{document}

\title{Generation of Flying Logical Qubits using Generalized Photon Subtraction \\ with Adaptive Gaussian Operations}

\author{Kan Takase}
%\email{takase@alice.t.u-tokyo.ac.jp}
\affiliation{\UTokyo}
\affiliation{\RQC}
\email{takase@ap.t.u-tokyo.ac.jp}

\author{Fumiya Hanamura}
\affiliation{\UTokyo}

\author{Hironari Nagayoshi}
\affiliation{\UTokyo}

\author{J. Eli Bourassa}
\affiliation{\Xanadu}

\author{Rafael N. Alexander}
\affiliation{\Xanadu}

\author{Akito Kawasaki}
\affiliation{\UTokyo}

\author{Warit Asavanant}
\affiliation{\UTokyo}
\affiliation{\RQC}

\author{Mamoru Endo}
\affiliation{\UTokyo}
\affiliation{\RQC}

\author{Akira Furusawa}
\email{akiraf@ap.t.u-tokyo.ac.jp}
\affiliation{\UTokyo}
\affiliation{\RQC}

\date{\today}

\begin{abstract}
The generation of a logical qubit called the Gottesman-Kitaev-Preskill qubit in an optical traveling wave is a major challenge for realizing large-scale universal fault-tolerant optical quantum computers. Recently, probabilistic generation of elementary GKP qubits has been demonstrated using photon number measurements and homodyne measurements. However, the generation rate is only a few Hz, and it will be difficult to generate fault-tolerant GKP qubits at a practical rate unless success probability is significantly improved. Here, we propose a method to efficiently synthesize GKP qubits from several quantum states by adaptive Gaussian operations. In the initial state preparation that utilizes photon number measurements, an adaptive operation allows any measurement outcome above a certain threshold to be considered as a success. This threshold is lowered by utilizing the generalized photon subtraction method. The initial states are synthesized into a GKP qubit by homodyne measurements and a subsequent adaptive operation. As a result, the single-shot success probability of generating fault-tolerant GKP qubits in a realistic scale system exceeds 10$\%$, which is one million times better than previous methods. This proposal will become a powerful tool for advancing optical quantum computers from the proof-of-principle stage to practical application.

\end{abstract}

%\pacs{03.67.-a,42.50.Dv,42.50.Ex}
% % 03.67.-a : Quantum information
% % 03.67.Hk : Quantum communication
% % 42.50.Dv : Quantum state engineering and measurements
% % 42.50.Ex : Optical implementations of quantum information processing and transfer

\maketitle

\section{Introduction}
Quantum computers are expected to demonstrate superior information processing capabilities compared to conventional computers by incorporating quantum phenomena such as superposition and entanglement. Applications of quantum computers include factoring, discrete optimization, quantum simulation, machine learning, and are anticipated to be used in fields such as finance, drug discovery, materials, and logistics \cite{Preskill2018quantumcomputingin}. Quantum computers that use optical traveling waves \cite{10.1063/1.5100160,Fukui_2022} can operate at room temperature and atmospheric pressure, achieving a clock frequency exceeding THz \cite{Ralph:99,Takanashi:20}. In addition to the quantum applications mentioned above, quantum computers based on optical traveling waves are well-suited for communication technology \cite{PhysRevA.58.146,PhysRevA.58.159}.

So, how far has the development of optical quantum computers progressed, and what are the challenges? Firstly, a scalable and programmable processor capable of performing measurement-based quantum gates has been experimentally demonstrated \cite{Asavanant2019,Larsen2019,Larsen2021}. One advantage of the optical approach is its ability to address scalability, a major challenge in other physical systems. While the clock frequency of the processors demonstrated so far is at most on the order of 10 MHz \cite{PhysRevApplied.16.034005}, there is a prospect of improvement to the order of 10 GHz through noiseless optical amplification technology \cite{10.1063/5.0137641}. However, the quantum advantage in information processing using these processors has not yet been demonstrated. A promising way to achieve practical quantum computation with universality and fault-tolerance is via Gottesman-Kitaev-Preskill (GKP) qubits \cite{Gottesman2001}. Once GKP Pauli eigenstates and magic states are prepared, only passive linear optics and homodyne measurement are needed \cite{Gottesman2001,PhysRevA.71.022316,PhysRevLett.123.200502,PhysRevResearch.2.023270,PRXQuantum.2.040353}. Making such advanced optical quantum states available is an urgent challenge for optical quantum computers.

GKP qubits belong to a class of states called non-Gaussian states. There are two typical methods for generating non-Gaussian states in propagating light. One is a method that utilizes the nonlinear interaction of light with matter or with light. The other is a method that utilizes quantum entanglement and measurements, known as the heralding method. In particular, one can prepare multimode entangled Gaussian states deterministically and employ non-Gaussian measurements such as photon number measurements on a subset of modes to herald non-Gaussian states on the remaining modes. In the former method, there is a problem that the nonlinearity of the optical traveling wave is inherently weak. Although methods like cavity quantum electrodynamics (QED) and quantum dots have achieved the generation of single photons \cite{Senellart2017} or Schrödinger cat states \cite{Hacker2019}, more advanced matter-light coupling and control are required for generating more complex states. Another challenging but interesting direction is converting non-Gaussian states generated in harmonic oscillators in other physical platforms such as trapped ions \cite{Flühmann2019} and superconducting circuits \cite{Campagne-Ibarcq2020}, where strong nonlinearity is easy to use, into optical traveling waves \cite{Mirhosseini2020,Holzgrafe:20}. However, methods that use matter tend to slow down the generation protocol, and they are not suitable for the high clock frequencies of optical quantum computers. In this regard, methods using light-light coupling have an advantage \cite{PhysRevLett.124.240503,PRXQuantum.4.010333}, but whether practical nonlinearities can be achieved awaits experimental demonstration. On the other hand, the heralding method is more feasible and widely used for generating non-Gaussian states. The heralding method is a non-unitary process conditioned by a measurement on a subsystem of entangled quantum light. Strong nonlinearity can be replaced with photon number measurements, enabling the generation of Schrödinger cat states \cite{Ourjoumtsev2006,Neergaard-Nielsen2006,Ourjoumtsev2007,PhysRevA.103.013710,PhysRevA.100.052301,anteneh2023machine,Eaton2022measurementbased}, arbitrary photon-number superposition states \cite{Yukawa:13,PhysRevApplied.15.024024,PhysRevA.100.052301}, and even elementary GKP qubits \cite{konno2023propagating,Takase2023,PRXQuantum.2.040353,PhysRevA.100.052301,PhysRevA.101.032315,Eaton2019}. These states can be generated in short pulses that are suitable for processors with a high clock frequency. However, a challenge of the heralding method is that the success of state generation is probabilistic, leading to low state generation rate. In the case of GKP qubits, the generation rate is only a few Hz even for the approximated GKP qubits with a low mean photon number \cite{konno2023propagating}. The most straightforward solution for circumventing this problem is to parallelize the state generation systems, but the number of parallel systems would likely become immense for quasi-deterministic state generation.

In this paper, we focus on the generation of GKP qubits and reduce the inherent probabilistic nature of the heralding method. Specifically, we introduce adaptive quantum operations into the Gaussian breeding protocol \cite{Takase2023}. The Gaussian breeding protocol entangles Gaussian states in a simple, iterative sequence of beam splitters and measures all-but-one mode with photon number measurements to `breed' the non-Gaussian measurement outcomes into approximate GKP qubits in the unmeasured mode. Gaussian breeding can generate GKP qubits with minimal resources and has a much higher success probability than the cat breeding protocol \cite{Vasconcelos2010,PhysRevA.97.022341} used in Ref. \cite{konno2023propagating}. Similarly to the Gaussian breeding protocol, the cat breeding protocol employs a simple, iterative sequence of beam splitters and measurements to combine input states into approximate GKP qubits; however, the input states are non-Gaussian (cat states), and the measurements are Gaussian (homodyne). We first reevaluate the advantages of Gaussian breeding and explicitly show the target wave function of state generation. This clear and concise explanation demonstrates why GKP qubits can be synthesized from photon number measurements, enabling efficient GKP qubit generation. The desired wave function can be synthesized by using generalized photon subtraction (GPS) \cite{PhysRevA.103.013710}. Furthermore, instead of considering only one measurement outcome as a success as in conventional heralding methods, our approach accepts multiple measurement outcomes and synthesizes desired states by performing adaptive operations based on the measurement outcomes. As a result, it is possible to generate fault-tolerant GKP qubits over a 10$\%$ probability by parallelizing only 20 GPS units, which contain 40 squeezed light sources and 20 photon-number-resolving detectors in total. Here, each squeezer needs to emit up to 18 dB squeezed vacuum states and each detector needs to resolve up to 20 photons. Compared to the original Gaussian breeding protocol with a system of similar size, the success probability is more than one million times better in the proposed protocol. This proposal will enable the generation of fault-tolerant GKP qubits at a practical rate in a realistic system, accelerating the realization of optical quantum computers.

%%%%%%%%%%%%%%%%% END OF INTRODUCTION %%%%%%%%%%%%%%%%%%%%%%%%%%%%%%%%%%%%%

%%%%%%%%%%%%%
\begin{figure}[t]
	\begin{center}
		\includegraphics[bb= 0 0 640 170,clip,width=\textwidth]{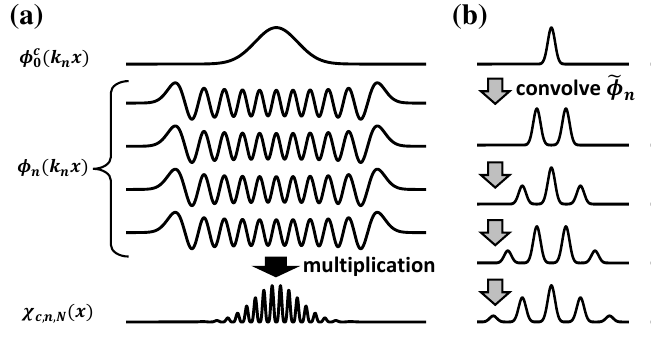}
	\end{center}
	\caption{Breeding protocol in a wavefunction picture. ${\bf (a)}$ Explanation in $x$ basis. The wave function of a squeezed vacuum state is the envelope, and the oscillation of $\phi_n(k_nx)$ create the comb structure. ${\bf (b)}$ Explanation in $p$ basis. By convolving $\tilde{\phi}_n$ to a Gaussian function iteratively, the number of superposed Gaussians increases.}
\label{Fig:1}
\end{figure}
%%%%%%%%%%%%

\section{METHOD}
\subsection{Target state\label{2a}} 
To achieve fault tolerance in quantum computing, quantum information must be robustly encoded as discrete variable states, such as qubits \cite{gottesman2009introduction}. In conventional qubit quantum error correction, many physical qubits are entangled to form a logical qubit \cite{PhysRevA.86.032324}. On the other hand, a harmonic oscillator can encode one logical qubit into one quantum mode \cite{PhysRevA.97.032346}. The GKP qubit is considered a promising logical qubit in harmonic oscillators. Once a GKP qubit is available, fault-tolerant universal quantum computations can be performed using only linear transformations of position and momentum operators, which are called Gaussian operations \cite{Gottesman2001,PhysRevA.71.022316,PhysRevLett.123.200502,PhysRevResearch.2.023270}.

Here we review basics of GKP qubits. Suppose that position and momentum operators of a harmonic oscillator satisfy the commutation relation $[\hat{x},\hat{p}]=i$. An ideal GKP qubit is defined as a superposition of equally spaced position eigenstates. In this paper, we consider states with $\sqrt{2\pi}$-spacing. Such a GKP qubit is also called qunaught state \cite{PhysRevA.102.062411} or sensor state \cite{PhysRevA.95.012305}. Such states can create Bell pairs with a beam splitter \cite{PhysRevA.102.062411}. The ideal GKP qubit is a non-physical state with infinite energy, so what we aim to generate is a state with the following wave function,
\begin{equation}\label{eq:sensor}
\varnothing_{\Delta_x,\Delta_p}^{(t)}(x) \propto \sum_{s=-\infty}^{\infty}e^{-\frac{\Delta_p^2}{2}\left((s+t/2)\sqrt{2\pi}\right)^2}e^{-\frac{1}{2\Delta_x^2}\left(x-(s+t/2)\sqrt{2\pi}\right)^2},\ t=0,1.
\end{equation}
$\varnothing_{\Delta_x,\Delta_p}^{(t)}(x)$ is a series of Gaussian functions of width $\Delta_x$ and separation $\sqrt{2\pi}$ embedded in a Gaussian envelope of width $1/\Delta_p$. When $\Delta_x, \Delta_p \rightarrow 0$, this state becomes an ideal GKP qubit. In order to perform fault-tolerant quantum computation, it is sufficient to achieve $\Delta_x<0.316$ (10 dB squeezing) \cite{Fukui2018,PRXQuantum.2.040353}. Suppose we have a GKP-like state with a density operator $\hat{\rho}$. The performance of this state as a GKP qubit is often evaluated by the effective squeezing \cite{PhysRevA.95.012305} of $x,p$ given by
\begin{equation}
\tilde{\Delta}_{x} = \sqrt{\frac{-1}{\pi}\ln{\left|{\rm Tr}\ e^{i\sqrt{2\pi}\hat{x}} \hat{\rho}\right|^2}},\ \tilde{\Delta}_{p} = \sqrt{\frac{-1}{\pi}\ln{\left|{\rm Tr}\ e^{-i\sqrt{2\pi}\hat{p}}\hat{\rho}\right|^2}}.
\end{equation}
In the case of Eq. (\ref{eq:sensor}), we get $\tilde{\Delta}_x=\Delta_x, \tilde{\Delta}_p=\Delta_p$. In this paper, we evaluate GKP-qubit generation using effective squeezing.

Finally, we introduce a function that approximates Eq. (\ref{eq:sensor}). We denote the wave function of $n$-photon states by $\phi_n(x)\propto H_n(x)e^{-\frac{1}{2}x^2}$, where $H_n(x)=(-1)^ne^{x^2}\frac{d^n}{dx^n}e^{-x^2}$. Then, the following relation is obtained,
\begin{equation}\label{eq:target}
e^{i\sqrt{\frac{\pi}{2}}sx}\varnothing_{\Delta_x,\Delta_p}^{(t)}(x) \approxprop \chi_{c,n,N}(x) \equiv \phi_0^c(k_nx)\cdot \phi_n^N(k_nx),
\end{equation}
where $\phi_i^{\nu}(x) \equiv \left(\phi_i(x)\right)^{\nu},  k_n = \sqrt{\frac{\pi}{2n+1}},s=N \ \mathrm{mod}\ 2$ and $t=n \ \mathrm{mod}\ 2$. This relation stems from the periodicity of $\phi_n(x)$. The Wentzel-Kramers-Brillouin approximation \cite{schleich2011quantum} shows that $\phi_n(x)$ has cosine (or sine) oscillation with an angular wavenumber $\sqrt{2n+1}$ when $|x| \ll 1$. Thus, $\phi_n^N(k_nx)$ has a comb structure with a spacing $\sqrt{2\pi}$. The term $\phi_0^c(k_nx)$ applies a Gaussian envelope and extracting only the comb structure. As a result, $\chi_{c,n,N}(x)$ closely approximates $\varnothing_{\Delta_x,\Delta_p}^{(t)}(x)$ as shown in Fig.\ref{Fig:1}(a). Considering this wave function in $p$-basis, GKP qubits can be approximated by convolution of the wave function of photon number states. Here, note that $\phi_n$ and its Fourier counterpart $\tilde{\phi}_n$ are the same function. The $p$-picture is shown in Fig.\ref{Fig:1}(b). The convolution of a Gaussian with appropriate width ($c \approx 1$) and a wave function of $n$-photon approximates a superposed Gaussians \cite{PhysRevA.103.013710}. By repeating convolution, a comb-shaped wave function is obtained. A similar principle is used in Gaussian breeding \cite{Takase2023}. Equation (\ref{eq:target}) can be considered as a concise expression of Gaussian breeding protocol.

%%%%%%%%%%%%%
\begin{figure*}[t]
	\begin{center}
		\includegraphics[bb= 0 0 510 360,clip,width=0.9\textwidth]{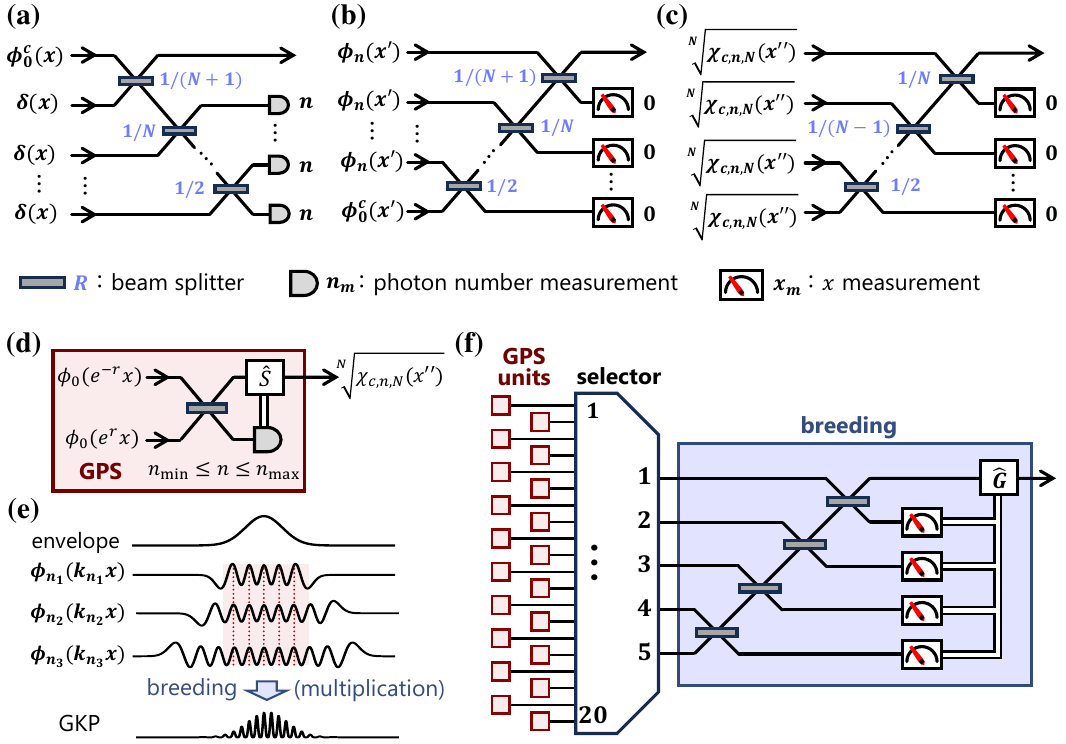}
	\end{center}
	\caption{How to realize the wave function $\chi_{c,n,N}(x)$. ${\bf (a)}$ Gaussian breeding. Photon number measurements are performed on multimode Gaussian states. $\delta(x)$ represents the infinitely squeezed state. ${\bf (b)}$ Inverted Gaussian breeding. Homodyne measurements are performed on Fock states. ${\bf (c)}$ A different form of the inverted Gaussian breeding. The input state $\sqrt[N]{\chi_{c,n,N}(x'')}=\phi_0^{c/N}(k_nx'')\cdot\phi_n(k_nx'')$ is an intermediate state of a Schrödinger cat state and a Fock state. ${\bf (d)}$ Generalized photon subtraction (GPS) to generate the state $\sqrt[N]{\chi_{c,n,N}(x'')}$. Adaptive squeezing is needed to incorporate a GPS circuit into the whole setup shown in (f). ${\bf (e)}$ Matching of the oscillation of different photon number states.  ${\bf (f)}$ Proposed system for GKP-qubit generation. A squeezing operation is performed on the output of GPS to generate an input state. $M$ GPS units are parallelized such that $N$ events that satisfy $n_{{\rm min}}\le n \le n_{{\rm max}}$ are selected. By inputting them to a breeding circuit, the desired wave function is realized. In the output, a Gaussian operation is performed according to the measured outcomes. Here, we show a case of $M=20,N=5$ which is discussed in the simulation.}
\label{Fig:2}
\end{figure*}
%%%%%%%%%%%%

\subsection{\textbf{Inverted Gaussian breeding} }

We will introduce three methods to realize the wave function $\chi_{c,n,N}(x)$ using the heralding scheme. The first is Gaussian breeding. If we set $\Delta_2=0,g=1,\Delta_3=1$ in Ref. \cite{Takase2023}, the desired wave function can be achieved with the setup shown in Fig. \ref{Fig:2}(a). The input states are a squeezed vacuum state $\phi_0^c(x)$ and infinitely squeezed vacuum states $\delta(x)$. The photon number measurements contribute to $\phi_n^N(x)$ component. On the other hand, the system in Fig. \ref{Fig:2}(b), which looks like a time reversed version of Fig. \ref{Fig:2}(a), similarly yields the output $\chi_{c,n,N}(x)$. Since homodyne conditioning has the effect of multiplicating input wave functions with squeezing, $\chi_{c,n,N}(x)$ can be synthesized by inputting a squeezed vacuum state $\phi_0^c\left(x'\right)$ and $N$ squeezed $n$-photon states $\phi_n\left(x'\right)$, where $x^{\prime}=\sqrt{N+1}x$. As a variant of Fig. \ref{Fig:2}(b), the desired wave function can also be obtained with Fig. \ref{Fig:2}(c), in which $\chi_{c,n,N}(x)$ is distributed evenly to each input as $\sqrt[N]{\chi_{c,n,N}\left(x''\right)}=\phi_0^{c/N}\left(k_nx''\right)\cdot\phi_n\left(k_nx''\right)$, where $x''=\sqrt{N}x$.

Figure \ref{Fig:2}(b) and (c) are similar to the cat breeding protocol \cite{Vasconcelos2010,PhysRevA.97.022341} in that homodyne conditioning is performed on non-Gaussian states. In fact, if we input Schrödinger cat states in the circuit of Fig. \ref{Fig:2}(b) and (c), we can realize cat breeding. However, if we compare the number of photons one needs to detect to prepare approximate cat states with GPS so that those states can be employed in the cat breeding protocol, with the number of photons one needs to detect in the Gaussian breeding protocol, we find that the Gaussian breeding protocol requires fewer photon detections to prepare an equivalent quality state, and hence makes better use of the non-Gaussian resources. In a similar vein, we find that the systems in Fig. \ref{Fig:2}(b) and (c) require fewer photons to be detected than the cat breeding protocol with GPS-produced cats. Compare these methods from the perspective of wave functions. From Ref. \cite{PhysRevA.103.013710}, the wave function of the cat state can be well approximated by $\phi_0 (k_n x)\cdot\phi_n (k_n x)$. Using this cat state, cat breeding outputs the wave function $\chi_{c=N,n,N}(x)=\phi_0^N(k_nx)\cdot\phi_n^N(k_nx)$. As $c$ becomes larger, the Gaussian envelope $\phi_0^c(k_nx)$ becomes sharper in Fig.\ref{Fig:1}(a), so the non-Gaussianity of $\phi_n^N (k_n x)$ is further impaired. Since $N$ is usually greater than 4, it is difficult to effectively utilize non-Gaussianity in cat breeding. On the other hand, the value of $c$ can be set freely in Fig. \ref{Fig:2}(a)-(c), and thus non-Gaussianity can be used efficiently. As an example, let us assume $n=20, N=5$. When $c=1$, the effective squeezing are $\tilde{\Delta}_x=0.34$ (9.3 dB) and $\tilde{\Delta}_p=0.19$ (14.3 dB). In cat breeding, we get $\tilde{\Delta}_x=0.34$ (9.4 dB) and $\tilde{\Delta}_p=0.33$ (7.2 dB). 

From the above discussion, Gaussian breeding and cat breeding can be uniformly explained using the wave function of Eq. (\ref{eq:target}). It is known that Gaussian breeding can efficiently synthesize GKP qubits, and the essence of this is that $c$ can be set freely. It is expected that the system shown in Fig. \ref{Fig:2}(b) and (c) will also be able to synthesize GKP qubits using a comparable number of photons detected as the Gaussian breeding protocol.

%%%%%%%%%%%%%%%%%%%%%%%%%%%%%%%%%%%%%%%%%%%%%%%%%%%%%%%%%%%%%%%%%%%%%%%%%%%%%
\subsection{Generalized photon subtraction}

The non-Gaussian input states of Fig. \ref{Fig:2}(b) and (c) can be generated by a heralding method called generalized photon subtraction (GPS) \cite{PhysRevA.103.013710}. GPS is originally a method of generating Schrödinger cat states in a two-mode Gaussian Boson Sampling setup \cite{PhysRevLett.119.170501}. The major difference between GPS and other similar approaches \cite{PhysRevA.101.032315,PhysRevA.100.052301} is that it introduces a wave function representation to express state generation analytically and comprehensively. Although we mainly discussed the generation of cat states expressed by $\left(\phi_0 * \phi_n\right)(x)\propto x^ne^{-x^2/4}$ in Ref. \cite{PhysRevA.103.013710}, GPS can generate various other states. Figure \ref{Fig:2}(d) is a setup of GPS. For now, ignore the squeeze operation in the output port of Fig. \ref{Fig:2}(d). For simplicity, we assume the wave functions of input squeezed vacuum states are given by $\phi_0(e^{-r} x)$ and $\phi_0(e^{r} x)$. When the transmittance of the beam splitter is $T=0.5$, we get a Einstein-Podolsky-Rosen (EPR) state. By detecting $n$ photons, we get $n$-photon state ($\phi_n(x)$). In the general case, the output state heralded by $n$-photon detection is given by
\begin{align}
\psi_n(x) &\propto \phi_0\left(x/\sqrt{a}\right)\cdot \left( \phi_0^{a}*\phi_n \right)(bx/a) \\
& \approxprop \phi_0^{e^{-2r}/T}\left(\sqrt{T/R}x\right)\cdot \phi_n\left(\sqrt{T/R}x\right) \ \ \ \ (e^{2r}\gg1),
\end{align}
where $a = Te^{-2r}+Re^{2r}, b = \sqrt{RT}(e^{-2r}-e^{2r}),R=1-T$ \cite{PhysRevA.103.013710}. Therefore, we can get $\phi_n(k_nx')$ or $\sqrt[N]{\chi_{c,n,N}(x'')}$ by squeezing the output of GPS depending on $n$.

%%%%%%%%%%%%%%%%%%%%%%%%%%%%%%%%%%%%%%%%%%%%%%%%%%%%%%%%%%%%%%%%%%%%%%%%%%%%%
\subsection{Incorporating adaptive elements}

The circuits in Fig. \ref{Fig:2}(a)-(c) are all heralding scheme, and the success probability with a single shot is $p\ll 1$. Therefore, when considering practical application, it will be necessary to add adaptive elements to these circuits so that the desired output can be obtained with higher probability. The simplest way is to parallelize the state generation system and exploit only successful events using a selector such as optical switches. A more advanced example is performing quantum operations in the output states depending on the heralding measurements. An example is the use of adaptive displacement in cat breeding to make the homodyne conditioning deterministic \cite{PhysRevA.97.022341}.

If adaptive elements are not taken into consideration, Fig. \ref{Fig:2}(a) would have the highest success probability and the highest feasibility among Fig. \ref{Fig:2}(a)-(c). As discussed in Ref. \cite{Takase2023}, Gaussian breeding shown in Fig. \ref{Fig:2}(a) has the advantage of minimal resource requirements and relatively high success probability, and is capable of generating meaningful states even at the current technological level. The resource requirements of Fig. \ref{Fig:2}(b) and (c) are higher than Fig. \ref{Fig:2}(a). When the input non-Gaussian states are generated using the heralding method, the cost just for generating the input states will be equal to or higher than Gaussian breeding. In addition, Fig. \ref{Fig:2}(b) and (c) requires conditioning by homodyne measurement, which further reduces the success probability. Strictly speaking, the probability of ideal conditioning with $x_m=0$ is 0.

So, which of Fig. \ref{Fig:2}(a), (b) and (c) is the most efficient when considering adaptive elements? Our conclusion is Fig. \ref{Fig:2}(c). The method in Fig. \ref{Fig:2}(c) is more compatible with adaptive elements than other methods for three reasons.

The first is that various photon detection patterns can be considered as a success. In Fig. \ref{Fig:2}(a), the ideal event would be for $N$ photon detectors to detect the same number of photons. As discussed in Ref. \cite{Takase2023}, multiple photon detection patterns such as $(n,n,n+2)$ or $(n-2,n,n+4)$ can also be considered as success, where a desired photon detection pattern is $(n,n,n)$. However, it is necessary that the number of detected photons does not differ too much and that the parity is the same. Also in Fig. \ref{Fig:2}(b) and (c), the basic success pattern is for $N$ GPS units to detect the same number of photons. In reality, this requirement can be significantly
relaxed relative to Fig. \ref{Fig:2}(a) by introducing adaptive elements. In our approach, to synthesize the wave function $\chi_{c,n,N}(x)$, it is essential to approximate the comb structure of GKP qubits with the oscillation of $\phi_n(x)$. Since the period and the phase of oscillation depends on $n$, squeezing and displacing the output state of GPS depending on $n$ helps to approximate $\phi_n^N(k_nx)$ efficiently. Explicitly, this relation is given by 
\begin{align}
\phi_0^c(k_nx)\cdot \phi_n^N(k_nx) &\approxprop \phi_0^c(k_nx)\cdot \prod_{s=1}^N \phi_{n_s}(k_{n_s}(x-d_s)),\\
n,n_s \ge n_{{\rm min}}, \ d_s &= \left\{
\begin{array}{ll}
0 & (n_s\ {\rm is\ even})\\
\sqrt{\frac{\pi}{2}} & (n_s\ {\rm is\ odd})
\end{array}
\right..
\end{align}
That is, in a region above a certain threshold $n_{{\rm min}}$, any photon detection event can be considered a success. This relation is visualized in Fig. \ref{Fig:2}(e). Importantly, such dynamic manipulation using different $n$ is possible in Fig. \ref{Fig:2}(b),(c), but not in Fig. \ref{Fig:2}(a). Furthermore, only a squeezing operation needs to be performed as a dynamic operation for the input non-Gaussian states because the displacement operation can be imposed on the subsequent homodyne conditioning. This is related to the advantage of optical GKP qubits: after generation, fault-tolerant quantum computation can be achieved by entangling GKP qubits into a cluster state that is only measured with homodyne detection \cite{PRXQuantum.2.040353,Bourassa2021blueprintscalable,PhysRevLett.112.120504,PRXQuantum.2.030325}.

Second, the requirement for simultaneity of photon detection events is reduced. In Fig. \ref{Fig:2}(a), the probability to detect $n$ photons in one mode is at most a few percent. Since this rare event needs to succeed $N$ times simultaneously, the overall success probability becomes quite low. In Fig. \ref{Fig:2}(b) and (c), it is necessary to generate $N$ non-Gaussian states by GPS, so the success probability of input state preparation would be about the same as Fig. \ref{Fig:2}(a). In Fig. \ref{Fig:2}(b) and (c), however, it is possible to utilize $M(> N)$ GPS units and select only $N$ successful events among them. There is no such way in Fig. \ref{Fig:2}(a) to avoid the simultaneous success of rare events. As $M$ increases, the probability of successful preparation of the input states improves dramatically.

Third, the strict homodyne conditioning with $x_m=0$ shown in Fig. \ref{Fig:2}(c) is not actually necessary. In addition to the ideal case of obtaining $x_m=0$ in all homodyne measurements, it is also possible to obtain a wave function close to $\chi_{c,n,N}(x)$ by performing adaptive Gaussian operations depending on the measurement results. This would be based on the same principle as the method of making homodyne conditioning deterministic in cat breeding \cite{PhysRevA.97.022341}, because the input state $\sqrt[N]{\chi_{c,n,N}(x'')}=\phi_0^{c/N}(k_nx'')\cdot\phi_n(k_nx'')$ is an intermediate state between a cat state $\phi_0(k_nx'')\cdot\phi_n(k_nx'')$ \cite{PhysRevA.103.013710} and a photon number state $\phi_n(k_nx'')$. On the other hand, the same method is not very effective in Fig. \ref{Fig:2}(b). A representative sample of the homodyne distribution in Fig. \ref{Fig:2}(b) originates from the side peaks of $\phi_n(k_nx)$ that are not essential in the concept shown in Fig. \ref{Fig:1}(a). Because the side peaks are suppressed  by the term $\phi_0^{c/N}(k_nx)$ in Fig. \ref{Fig:2}(c), there are many more homodyne outcomes that prepare comparable states to that created by homodyne outcome 0.

Based on the above discussion, we propose Fig. \ref{Fig:2}(f), which is an adaptive version of Fig. \ref{Fig:2}(c), as a GKP qubit generation system. The input state $\phi_0^{c/N}(k_nx'')\cdot\phi_n(k_nx'')$ is generated by GPS. The output of GPS is squeezed according to the measured photon number. State preparation with GPS is considered as success when the measured photon number is $n_{{\rm min}}\le n\le n_{{\rm max}}$. Note that we set a maximum photon detection number $n_{{\rm max}}$ to simulate actual experimental requirement. $M$ GPS units are spatially parallelized, and a selector consisting of optical switches selects $N$ successful events and inputs them into a breeding circuit. An adaptive Gaussian operation is performed on the output state of the breeding circuit according to the measurement results. In the next section, we will simulate GKP qubit generation in this system and evaluate its performance.

\subsection{Simulation\label{2d}}

\begin{table}[hbtp]
  \caption{Simulation results with $N=M=5$. input squeezing: squeezing level of GPS inputs, $P_{{\rm NGS}}$: probability to detect $n$ photons between $n_{{\rm min}}$ and $n_{{\rm max}}$ in each GPS unit, $P_{{\rm HD}}$: success probability of homodyne conditioning when $N$ non-Gaussian inputs are prepared, $P_{{\rm total}}$: success probability of whole system given by $P_{{\rm NGS}}^N\cdot P_{{\rm HD}}$.}
  \label{table:1}
  \centering
  \begin{tabular}{|c|c|c|c|c|c|c|}
    \hline
    $\ \ $$n_{{\rm min}}$$\ \ $ & $\ \ $$n_{{\rm max}}$$\ \ $ & $ c $&$\ \ $input squeezing$\ \ $ & $\ \ P_{{\rm NGS}}\ \ $& $\ \ P_{{\rm HD}}\ \ $ & $\ \ P_{{\rm total}}\ \ $ \\
    \hline
    10 & 20 & 1.3& 17.7 dB  & 19$\%$ & 30$\%$ & 0.0080$\%$ \\
    10 & 30 & 1.4& 18.7 dB  & 28$\%$ & 40$\%$ & 0.072$\%$  \\
    10 & 40 & 1.4& 19.4 dB  & 34$\%$ & 47$\%$ & 0.23$\%$ \\
    \hline
    6 & 20 & 1.3& 17.7 dB  & 32$\%$ & 13$\%$ & 0.049$\%$ \\
    6 & 30 & 1.4& 18.7 dB  & 41$\%$ & 18$\%$ & 0.21$\%$  \\
    6 & 40 & 1.4& 19.4 dB  & 46$\%$ & 26$\%$ & 0.54$\%$ \\
   \hline
  \end{tabular}
\end{table}

%%%%%%%%%%%%%
\begin{figure*}[t]
	\begin{center}
		\includegraphics[bb= 0 0 480 330,clip,width=0.8\textwidth]{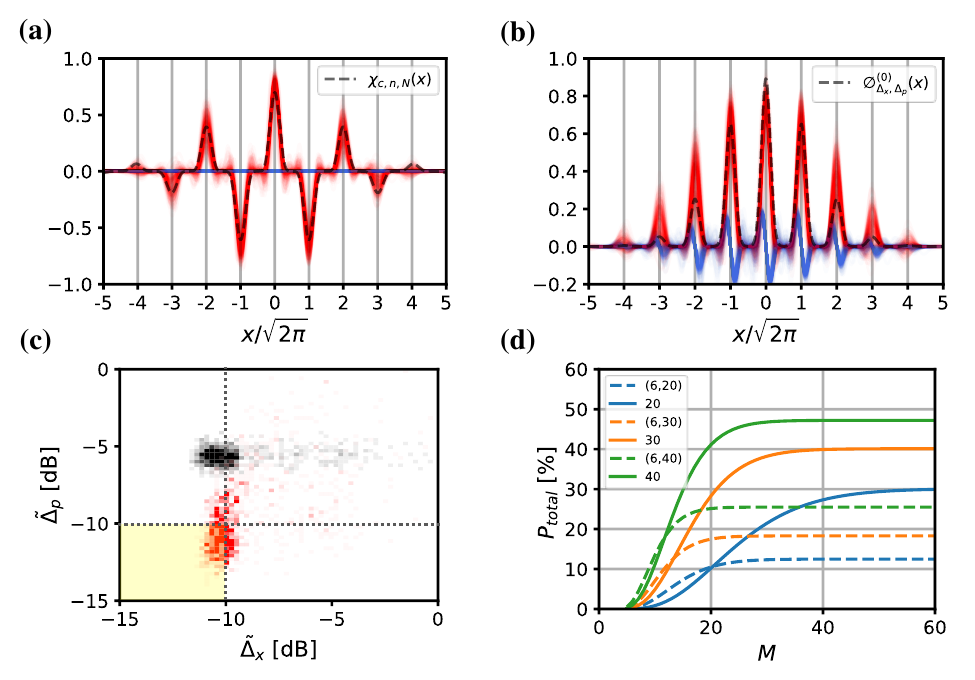}
	\end{center}
	\caption{Simulation results. ${\bf (a)}$ Overlay of the wave functions of successful events with $c=1.3, n_{{\rm min}}=10, n_{{\rm max}}=20, N=5$. The red and blue lines are the real and imaginary components, respectively. The target state (black dashed line) is $\chi_{c,n=20,N}(x)$. ${\bf (b)}$ Overlay in the same condition with (a), where the target state is a GKP qubit with 10 dB effective squeezing. ${\bf (c)}$ Distribution of effective squeezing. Red: proposed protocol with $c=1.3, n_{{\rm min}}=10, n_{{\rm max}}=20,N=5$. Black: cat breeding in the same condition, that is, $c=5, n_{{\rm min}}=10, n_{{\rm max}}=20,N=5$. ${\bf (d)}$ $M$-dependence of the total success probability under the conditions shown in Table \ref{table:1}. The legend in the plot shows $(n_{{\rm min}},n_{{\rm max}})$. }
\label{Fig:3}
\end{figure*}
%%%%%%%%%%%%

We evaluate the proposed method by Monte Carlo simulation by using Mr Mustard, a numerical simulation and optimization Python package for quantum optics \cite{MrMustard}. The state generation is considered as a success when effective squeezing $\tilde{\Delta}_x,\tilde{\Delta}_p$ is 10 dB or more. First, we consider the case where minimal GPS units are used ($N=M$). Table \ref{table:1} shows simulation results for various $n_{{\rm min}}$ and $n_{{\rm max}}$ when $N=M=5$. The value of $c$ is optimized to maximize the total success success probability. The input squeezing level for GPS circuits is chosen so that $P_{{\rm NGS}}$, the probability to detect photons between $n_{{\rm min}}$ and $n_{{\rm max}}$, is maximized. $P_{{\rm HD}}$ is the success probability of homodyne conditioning when $N$ desired non-Gaussian states are supplied. A Gaussian operation is performed to optimize effective squeezing for each output state. The total success probability is given by $P_{{\rm total}}=P_{{\rm NGS}}^N\cdot P_{{\rm HD}}$.

In Table \ref{table:1}, $P_{{\rm NGS}}$ are all above 10$\%$. This is a clear advantage over previous methods, which have at most a few percent probability to detect $n$ photons. Our approach improves this problem by utilizing a wide range of photon detection events, leading to significant increase of the total success probability. $P_{{\rm HD}}$ is also a high value over 10$\%$. $P_{{\rm total}}$ is about $0.01\%$ to $0.5\%$. This is a high probability as a single-shot state generation of GKP qubits. For example, the probability for generating equivalent states with Gaussian breeding is roughly estimated to be $\left(2\cdot10^{-2}\right)^N\approx 10^{-6}\%$. The adaptive elements improve the probability more than one thousand times. Figure \ref{Fig:3}(a) shows overlayed wave functions of the output states of successful events when the target state is $\chi_{c,n,N}(x)$. Figure \ref{Fig:3}(b) shows a similar plot when the target state is $\varnothing_{\Delta_x,\Delta_p}^{(0)}(x)$. Although there are differences in peak height for each event, they all well-approximate the target states. The wave functions in Fig. \ref{Fig:3}(b) have imaginary component because they are obtained by displacing the wave functions in Fig. \ref{Fig:3}(a) in $p$ direction as shown in Eq. (\ref{eq:sensor}). The ratio of the imaginary component is about 7$\%$. Note that this imaginary component does not appear when $N$ is even.

Next, we compare our protocol with cat breeding. Figure \ref{Fig:3}(c) is $\tilde{\Delta}_x,\tilde{\Delta}_p$ distribution of the simulation results. The red plot is the result of the method we proposed, which yields approximately 10 dB of effective squeezing both in position and momentum. The black plot is the result of cat breeding using approximate cats produced from GPS ($c=N$), which clearly results in less squeezing. To get 10 dB effective squeezing in cat breeding, we need to set $n_{{\rm max}}$ much larger. This shows that our method can efficiently generate GKP qubits from fewer photon detections.

Finally, let us consider increasing $M$. The total probability is given by
\begin{equation}
P_{{\rm total}} = P_{{\rm HD}}\cdot\sum_{j=N}^M \ {}_M \mathrm{C}_j P_{{\rm NGS}}^j\cdot (1-P_{{\rm NGS}})^{M-j}
\end{equation}
As shown in Fig. \ref{Fig:3}(d), as $M$ increases, $P_{{\rm total}}$ increases rapidly and eventually approaches $P_{{\rm HD}}$. The optimal value for $n_{{\rm min}}$ also changes depending on how large $M$ can be. From Table \ref{table:1}, the smaller $n_{{\rm min}}$ is, the larger $P_{{\rm NGS}}$ becomes, but the smaller $P_{{\rm HD}}$ becomes. Therefore, if $M$ is small, it is advantageous to make $n_{{\rm min}}$ small. Let us set $P_{{\rm total}}=10\%$ as one goal. In this case, it can be realized by $(n_{{\rm min}}, n_{{\rm max}},M)=(10,20,20),(10,30,13),(10,40,10)$. The system with $M=20,N=5$ is shown in Fig. \ref{Fig:2}(f). Therefore, if we can use 18 to 20 dB squeezed vacuum states, photon number resolving capability from 20 to 40, and adaptive operations, it is possible to generate practical GKP qubits with a realistic system scale. Assuming that 20 photon-number-resolving detectors are available, let us estimate the success probability of Gaussian breeding. Since $N=5$, four circuits of Fig. \ref{Fig:2}(a) can be parallelized. The probability to obtain at least one GKP qubits out of the four circuit is about $4\cdot 10^{-6}\%<10^{-5}\%$. Therefore, our protocol can improve the total success probability more than one million times compared to previous methods.

\section{DISCUSSION}

We discussed how to overcome the probabilistic nature of heralding methods using adaptive elements, focusing on the GKP qubit. Our proposal will be a powerful technique for generating fault-tolerant GKP qubits at a practical rate. In order to realize our protocol, the following four points will be important.

First, it is necessary to verify how much photon loss deteriorates the generated states. In Ref. \cite{Takase2023}, the robustness of Gaussian breeding against photon loss has been discussed. The performance of our protocol in the presence of photon loss is the subject of future work \cite{xanaduprep}. Additionally, we should attempt to reduce both $n_{{\rm max}}$ and the input squeezing level, as the impact of loss becomes more pronounced when dealing with many photons. Introducing displacements before the photon number measurement might lower the squeezing requirement below the current world record of 15 dB \cite{PhysRevLett.117.110801}.

The second aspect involves the development of adaptive quantum operations. Both linear \cite{PhysRevLett.113.013601} and nonlinear \cite{Sakaguchi2023} feedforward, depending on homodyne measurements, have already been demonstrated as core technologies for optical quantum information processors. However, adaptive operations based on photon number measurements have not yet been realized and need further development. Also, it is necessary to consider which adaptive operations are suitable to implement the proposed method. For example, instead of squeezing the GPS outputs, the reflectance of beam splitters in the breeding circuit can be adaptively changed \cite{xanaduprep}.

The third aspect is improvements of architecture. While it may be straightforward to use ordinary optical switches as the selector, it is also promising to explore mode selection through quantum teleportation \cite{PhysRevA.107.032412}. For the preparation of input states, leveraging timing synchronization via quantum memories has proven to be powerful \cite{PhysRevLett.110.133601}. This can be interpreted as multiplexing of GPS units in the time domain. Additionally, frequency domain multiplexing \cite{Joshi2018}, taking advantage of the broadband nature of light, is a viable option. Through the use of such multiplexing, apart from spatial parallelization, the proposed method can be implemented more efficiently. Concerning the overall system architecture, alternatives such as two-dimensional cluster states \cite{Asavanant2019,Larsen2019,Larsen2021} or a loop-based quantum processor \cite{doi:10.1126/sciadv.abj6624,PhysRevLett.131.040601} can be considered instead of the interferometer configuration shown in Fig. \ref{Fig:2}(f). In other words, an optical quantum computer, which is a general-purpose quantum manipulation platform, is also useful as a quantum state synthesizer.

Finally, it is important to develop heralding state generation using multiphoton detection. In heralding methods, quantum non-Gaussianity could have been confirmed using up to three-photon detection \cite{Gerrits2010,Yukawa:13,Endo:23}. However, to realize the proposed method, it is essential to generate quantum state using photon detection with more than 10 photons. Bridging this significant technological gap requires the development of various techniques, such as high-performance photon-number-resolving detectors capable of detecting more than 10 photons\cite{10.1063/5.0149478}, a mode-selective heralding method \cite{Ra2020}, domain engineering in parametric down-conversion sources \cite{Pickston:21}, and low-loss state evaluation using noiseless optical amplification \cite{10.1063/5.0137641}.

By addressing the above points, we can eliminate the bottleneck in the development of optical quantum computers.

%%%%%%%%%%%%%%%%%%%%%%%%%%%%%%%%%%%%%%%%%%%%%%%%%%%%%%%%%%%%%%%%%%%%%%%%%%%%

\section{Acknowledgements}
This work was partly supported by Japan Science and Technology (JST) Agency (No. JPMJMS2064), UTokyo Foundation, and donations from Nichia Corporation of Japan. K.T., F.H., and W.A. acknowledge the funding from Japan Society for the Promotion of Science KAKENHI (No.22K20351, 23K13038, 23KJ0498, 23K13040). M. E. acknowledges the funding from JST (JPMJPR2254). K.T, W.A. and M.E. acknowledge supports from Research Foundation for OptoScience and Technology. F.H., H.N., and A.K. acknowledge support from FoPM, WINGS Program, the University of Tokyo. A.K. acknowledges financial support from Leadership Development Program for Ph.D. (LDPP), the University of Tokyo. J.E.B. and R.N.A. acknowledge the contributions of  Rachel Chadwick, Sebastian Duque-Mesa, Scott Glancy, Jacob Hastrup, Timo Hillmann, Carlos Lopetegui, Filippo Miatto, Zeyue Niu, Yong-Siah Teo, Ilan Tzitrin, Yuan Yao.

%%%%%%%%%%%%%%%%%%%%%%%%%%%%%%%%%%%%%%%%%%%%%%%%%%%%%%%%%%%%%%%%%%%%%%%%%%%%%%%%%%%%%%%%%%%%%%%%%%%%%%%%%%%%

\bibliographystyle{apsrev4-1}

%unsrt

%\bibliography{scibib}


\end{document}